\documentclass{arxiv} 

\newif\ifarxiv
\arxivtrue

\usepackage{rotating}
\usepackage{nicematrix}
\usepackage{pgfplotstable}

\usetikzlibrary{external}
\tikzexternaldisable

\usetikzlibrary{patterns,calc,tikzmark,decorations,pgfplots.groupplots,intersections}
\usepgfplotslibrary{fillbetween}
\pgfplotsset{
    every crossref picture/.style = {
        baseline,
        yshift = 0.45ex
    },
    bluestyle/.style ={
        blue,
        thick
    },
    redstyle/.style ={
        red,
        dashed,
        thick
    },
    yellowstyle/.style ={
        yellow,
        dash dot,
        thick
    },
    greenstyle/.style ={
        green,
        dash dot dot,
        thick
    }
}

\newcounter{step}
\newcommand{\step}[1]{\subsection*{Step~\ctext{\thestep} \hspace{0.5em} #1} \stepcounter{step}}
\newcommand{\stepref}[1]{Step~\ctext{#1}}

\newlength{\cu}
\colorlet{greenfill}{green!30}
\colorlet{bluefill}{blue!30}
\definecolor{string}{RGB}{170,0,255}

\newcommand{\mpoly}[1][\phantom{j}]{\ensuremath{\mcl{P}_{\negthickspace #1}}}
\newcommand{\poly}[1][\phantom{j}]{\ensuremath{p_{#1}}}
\newcommand{\codetilde}{\textasciitilde}
\newcommand{\displaylabel}[1]{(~\protect\ref{#1}~)}
\newcommand{\ctext}[1]{\tikz[baseline = -0.625ex]{\node[circle,draw,inner sep = 0.5pt,font = \bfseries] at (0,0) {#1};}}

\title{Solving Multivariate Polynomial Systems and Rectangular Multiparameter Eigenvalue Problems with MacaulayLab\thanks{This work was supported in part by the KU Leuven Research Fund (grants IBOF/23/064, C3/20/117, and C3I/21/00316); in part by the FWO (grants S005319 and T001919N); in part by the Departement Economie, Wetenschap \& Innovatie via the Flanders AI Research Program; in part by the Vlaams Agentschap Innoveren \& Ondernemen (grant HBC/2021/0076); and in part by the European Research Council (grant 885682). \textit{Views and opinions expressed are, however, those of the authors only and do not necessarily reflect those of the European Union or ERC. Neither the European Union nor the granting authority can be held responsible for them.}}}

\author[$\dagger$, $\ddagger$]{Christof Vermeersch}
\author[$\ddagger$]{Bart De Moor}

\affil[$\dagger$]{Corresponding author (\url{christof.vermeersch@esat.kuleuven.be})}
\affil[$\ddagger$]{Center for Dynamical Systems, Signal Processing, and Data Analytics (STADIUS), Dept. of Electrical Engineering (ESAT), KU Leuven, Kasteelpark Arenberg 10, 3001 Leuven, Belgium}

\begin{document}
    \maketitle
    
    \ifarxiv
    \begin{abstract}
        We present the \matlab toolbox \macaulaylab, which implements numerical linear algebra algorithms for solving multivariate polynomial systems and rectangular multiparameter eigenvalue problems. 
        Its structure and functionality are the result of several years of research and algorithmic development.
        We demonstrate how the software works and compare its performance with other software packages, such as \toolbox{PNLA}, \toolbox{PHCpack}, and \toolbox{MultiParEig}.
        Some core features of \macaulaylab are the fact that it solves two key problems via one common approach, works independently of the chosen polynomial basis and monomial order, and is capable of dealing with positive-dimensional solution sets at infinity.
        The toolbox (including its future updates) and a large collection of test problems are freely available online.
    \end{abstract}
\fi

\section{Introduction}
    \label{sec:introduction}

    \macaulaylab is a \matlab toolbox that uses numerical linear algebra techniques to solve two challenging, but important, problems---it tackles multivariate polynomial systems and rectangular multiparameter eigenvalue problems.
    On the one hand, finding the common roots of multivariate polynomial systems is a well-known problem in algebraic geometry and computational nonlinear algebra. 
    Because of the omnipresence of multivariate polynomials in scientific models and engineering applications, there have been put a lot of effort in developing solvers for multivariate polynomial systems.
    While iterative root-finding methods, such as the efficient homotopy continuation methods, track solution paths, algebraic root-finding methods transform the system into (multiple) univariate root-finding or eigenvalue problems. 
    The algorithms implemented in \macaulaylab take the latter approach to compute the common roots and produce one or multiple generalized eigenvalue problems.
    Rectangular multiparameter eigenvalue problems, on the other hand, are less known for the general audience, but also appear in various modeling efforts.
    \macaulaylab is one of the first toolboxes focussing on rectangular multiparameter eigenvalue problems. 
    Similar as for the polynomials, the algorithms in \macaulaylab rephrase the rectangular multiparameter eigenvalue problems as one or multiple generalized eigenvalue problems.
    Very recently, the authors of~\cite{hochstenbach2024numerical} have shown that it is also possible to translate (linear) rectangular problems into a more established square formulation, enabling a new set of algorithms to be used.

    While these two key problems may seem unrelated at first glance, multivariate polynomial systems and rectangular multiparameter eigenvalue problems are both special cases of a broader problem and are connected through the (block) Macaulay matrix.
    \Cref{sec:problemdefinition} gives a rigorous definition of both key problems and reveals their relation.
    In that sense, \macaulaylab is quite unique; the combined nature of the two key problems allows for one tool(box) to tackle them both.
    
    \macaulaylab uses the (block) Macaulay matrix to solve multivariate polynomial systems and rectangular multiparameter eigenvalue problems via a similar numerical linear algebra methodology. 
    As explained in \cref{sec:blockmacaulaymatrix}, it transforms both key problems into a multidimensional realization problem by using the column space or (right) null space of the (block) Macaulay matrix constructed from the coefficients of the polynomials or the coefficient matrices of the rectangular multiparameter eigenvalue problem. 
    The multidimensional realization problem corresponds to one or multiple generalized eigenvalue problems that yield the solutions of the original problem.
    We explain in \cref{sec:structure} how this transformation is obtained by only using numerical linear algebra tools, such as singular value, QR, or eigenvalue decompositions.
    The toolbox relies on the decades of advancements in numerical linear algebra, resulting in computationally efficient and numerically robust algorithms.
    An important feature of \macaulaylab is that all the algorithms are implemented without depending on a particular polynomial basis or monomial order, allowing the user to choose what suits the application best (see \cref{sec:basis&order}).
    It is known that in some situations an orthogonal polynomial basis, like the Chebyshev polynomials, has superior numerical properties~\cite{trefethen2019approximation,trefethen2011six}, while results in algebraic geometry often depend on the chosen monomial order~\cite{faugere1993efficient}.
    Offering the user the choice to select a particular polynomial basis and monomial order can thus be very useful in applications.

    \macaulaylab is written in \matlab and freely available at~\cite{vermeersch2025website}.
    Future updated versions of the software will be made available on the same website.
    The toolbox has been developed to be user-friendly and easy to use, whether the user simply wants to solve a problem or wants to learn more about its properties.
    Furthermore, the authors have gathered an extensive collection of test problems (see \cref{sec:database} and~\cite{vermeersch2025website}) and made these problems available together with the toolbox.  
    These problems can be used to experiment with \macaulaylab, but also to test other software packages, as illustrated in the comparison of \cref{sec:comparison}. 

    \paragraph*{Summary of the paper's outline and objectives}

        We continue this paper by rigorously defining the key problems that \macaulaylab tackles in \cref{sec:problemdefinition}.
        We explain why we discuss both multivariate polynomial systems and rectangular multiparameter eigenvalue problems and give a summary of the existing solution software.
        The common (block) Macaulay matrix approach is discussed in \cref{sec:blockmacaulaymatrix}.
        \emph{The paper's first objective is to discuss the implementation of the toolbox and present its most important functionalities.}\footnote{Notice that the paper is not a guide on how to use \macaulaylab. For more information about using the toolbox, we refer the interested reader to the user manual and introductory \matlab script (also available online at~\cite{vermeersch2025website}.}, which is done in \cref{sec:structure}. 
        Next, we highlight one of the most important features in \cref{sec:basis&order}: the independence of the polynomial basis and monomial order.
        \Cref{sec:functionalities} compares some of the solution strategies in \macaulaylab, which is explained by means of the accompanying collection of test problems.
        \emph{A second objective is to evaluate the software's performance with respect to other state-of-the-art solvers}, which is the topic of \cref{sec:comparison}.
        With that in mind, we discuss the current limitations and sketch the path of future developments in \cref{sec:limitations&developments} and conclude the paper in \cref{sec:conclusion}.

    \paragraph*{Hardware specifications}
    
        All computations in this paper are performed on a MacBook Pro that has an M1 CPU (2020) working at $\SI{3.2}{\GHz}$ (8 cores) and $\SI{16}{\giga\byte}$ RAM. 
        The operating system is macOS Sequoia and the \matlab version is R2024a.
        This choice has been made to evaluate the performance and capabilities of the toolbox under typical working conditions, rather than on a server with extensive memory resources.

        
\section{Two key problems and the software to solve them}
    \label{sec:problemdefinition}

    Let $\mpoly[j](\vc{x})$ be a matrix polynomial in $m$ variables, $\vc{x} = (x_1, \ldots, x_m)$, with degree $d_j$ and $k \times l$ coefficient matrices $\mt{A}^{(\vc{i})}_{j} \in \mathbb{C}^{k \times l}$.
    For example, the cubic matrix polynomial
    \begin{equation}
        \mpoly[1](\vc{x}) = \mt{A}^{(13)}_1 x_1 x_2^2 + \mt{A}^{(00)}_1 = 
        \begin{bmatrix}
            x_1 x_2^3 + 1 & 2 x_1 x_2^3 + 2 \\ 
            3 x_1 x_2^3 + 3 & 4 x_1 x_2^3 + 4 \\ 
            5 x_1 x_2^3 + 5 & 6 x_1 x_2^3 + 6
        \end{bmatrix},
        \quad \text{with} \, \mt{A}^{(13)}_1 = \mt{A}^{(00)}_1 = 
        \begin{bmatrix}
            1 & 2 \\ 
            3 & 4 \\ 
            5 & 6 
        \end{bmatrix}.
    \end{equation}
    The subscript indicates to which matrix polynomial the coefficient matrix belongs, while the multi-index $\vc{i} = (i_1, \ldots, i_n) \in \mathbb{N}^m$ in the superscript indexes the coefficient matrices and corresponds to the power of the associated monomial $\vc{x}^{\vc{i}}$.
    Consider now the system of $s$ matrix equations in $m$ variables,
    \begin{equation}
        \label{eq:unifyingproblem}
        \mpoly[1](\vc{x}) \, \vc{y} = \cdots = \mpoly[s](\vc{x}) \, \vc{y} = \vc{0}, \quad \left\Vert \vc{y} \right\Vert = 1,
    \end{equation}
    in which each matrix polynomial is multiplied by the same nonzero vector $\vc{y} \in \mathbb{C}^{l \times 1}_0$.
    \macaulaylab can be used to find the affine zero-dimensional solutions $(\vc{x}^\ast, \vc{y}^\ast)$ that satisfy these matrix equations.
    Two well-known key problems can be distilled from~\eqref{eq:unifyingproblem}: the multivariate polynomial root-finding problem (\cref{sec:polynomialdefinition}) and the rectangular multiparameter eigenvalue-finding problem (\cref{sec:eigenvalueproblem}).
    
    \subsection{Multivariate polynomial root-finding}
        \label{sec:polynomialdefinition}
    
        Suppose that the coefficient matrices are scalars, i.e., $k = l = 1$ such that $\mt{A}^{(\vc{i})}_j = a^{(\vc{i})}_j \in \mathbb{C}$. 
        Consequently, $\vc{y} \in \mathbb{C}^{1 \times 1}$ is a scalar with $\left\Vert \vc{y} \right\Vert = 1$; it is always equal to one and can thus be omitted.
        The $s$ matrix polynomials are multivariate polynomials for the first key problem.

        \subsubsection{Multivariate polynomial root-finding problem}

            Multivariate polynomials are powerful tools to model problems from various origins. 
            Finding the common roots of a multivariate polynomial system can be defined as follows.

            \begin{definition}
                \label{def:polynomialsystem}
                Given $s$ polynomials $\poly[j](\vc{x})$ with coefficients $a^{(\vc{i})}_j \in \mathbb{C}$, the \emph{multivariate polynomial root-finding problem} consists in finding all $m$-tuples $\vc{x}^\ast \in \mathbb{C}^m$, so that each $\poly[j](\vc{x}^\ast) = 0$, with $j = 1, \dots, s$.
            \end{definition}

            \begin{example}
                Consider the following system of two bivariate polynomials of degree two:
                \begin{equation}
                    \left\lbrace
                    \begin{aligned}
                        p_1 (x_1, x_2) &= a^{(20)}_1 x_1^2 + a^{(11)}_1 x_1 x_2 + a^{(02)}_1 x_2^2 + a^{(10)}_1 x_1 + a^{(01)}_1 x_2 + a^{(00)}_1, \\
                        p_2 (x_1, x_2) &= a^{(20)}_2 x_1^2 + a^{(11)}_2 x_1 x_2 + a^{(02)}_2 x_2^2 + a^{(10)}_2 x_1 + a^{(01)}_2 x_2 + a^{(00)}_2. 
                    \end{aligned}
                    \right.
                \end{equation}
                It is clear that $\vc{y}$ in the multivariate polynomial system is equal to one.
            \end{example}

            Because of their modeling capabilities, multivariate polynomial systems have applications in diverse areas of science and engineering, for example, in robotics, game theory, computational chemistry and biology, computer vision , system identification, and model order reduction. 
            Extensive lists of possible applications can be found in reference books like~\cite{cox2020applications,morgan1987solving,sommese2005numerical}.

        \subsubsection{Existing root-finding software}

            Currently, the most efficient way of obtaining the common roots of a system of multivariate polynomials with available software is via homotopy continuation algorithms.
            These algorithms employ a mixture of techniques from algebraic geometry and nonlinear optimization to continuously deform a starting system with known solutions into the original system with unknown solutions, while tracking the paths of the solutions (see, for example,~\cite{li1997numerical,verschelde1996homotopy,sommese2001numerical}). 
            Although issues with ill-conditioning still exist, homotopy continuation methods are inherently parallel, i.e., each isolated solution can be computed independently, and are currently among the most competitive algorithms to solve systems of multivariate polynomial equations.
            Their main disadvantage is that they only work for square (i.e., the number of equations is equal to the number of unknown variables) systems of multivariate polynomial equations.
            Because of their efficiency and applicability, software for homotopy continuation comes in many flavours.
            Some toolboxes that use homotopy continuation to tackle polynomial systems are \href{http://homultiparameter eigenvalue problemages.math.uic.edu/~jan/PHCpack/phcpack.html}{\toolbox{PHCpack}}~\cite{verschelde1999algorithm}, \href{https://people.cs.vt.edu/~ltw/hompack/hompack90.html}{\toolbox{HOMPACK}}~\cite{watson1987algorithm}, \toolbox{PHoM}~\cite{gunji2004phom,gunji2006phompara}, \href{http://www.hom4ps3.org}{\toolbox{HOM4PS}}~\cite{chen2014hom4ps}, \href{https://antonleykin.math.gatech.edu/NAG4M2}{\toolbox{NAG4M2}} as part of \href{https://faculty.math.illinois.edu/Macaulay2/}{\toolbox{Macaulay2}}~\cite{leykin2011numerical}, \href{https://bertini.nd.edu}{\toolbox{Bertini}}~\cite{bates2013bertini}, and \href{https://www.juliahomotopycontinuation.org}{\toolbox{HomotopyContinuation.jl}}~\cite{breiding2018homotopy}.

            Another approach to deal with systems of multivariate polynomial equations is to reduce the problem into subsequent univariate root-finding problems or eigenvalue problems.
            On the one hand, symbolic software packages, like \toolbox{Maple}, \toolbox{Singular}, \toolbox{Magma}, and \toolbox{msolve}, use a symbolic approach to solve polynomial systems, by creating a Gr\"{o}bner basis to create a triangular system that can be solved via back-substitution or to construct the multiplication matrices (the eigenvalues of which are related to the solutions of the polynomial system).
            It is, on the other hand, also possible to rely solely on numerical linear algebra techniques to rephrase polynomial systems as eigenvalue problems. 
            \macaulaylab falls into this category of solvers.
            Dreesen~\cite{dreesen2013back} and Batselier~\cite{batselier2013numerical} have also approached the problem via the Macaulay matrix and have implemented their algorithms in the \matlab packages \toolbox{RootFinding}\footnote{The \matlab package \toolbox{RootFinding} is available upon request with its developer, Philippe Dreesen.} and \href{https://github.com/kbatseli/PNLA_MATLAB_OCTAVE}{\toolbox{PNLA}}, respectively.
            Similar to \macaulaylab, they both use the Macaulay matrix to set up one or multiple multidimensional realization problems.
            While \toolbox{RootFinding} is a quite naive implementation of the Macaulay matrix approach as described in~\cite{dreesen2013back}, \toolbox{PNLA} offers functions that take advantage of the structure and sparsity of the Macaulay matrix.
            When the system consists of two bivariate polynomials, the \matlab package \href{https://www.mathworks.com/matlabcentral/fileexchange/54159-biroots}{\toolbox{BiRoots}} by~\cite{plestenjak2016roots} can be used, which transforms the problem into a square multiparameter eigenvalue problem and solves this problem via \href{https://nl.mathworks.com/matlabcentral/fileexchange/47844-multipareig}{\toolbox{MultiParEig}} (see \cref{sec:eigenvalueproblem}).
            For users of the Julia language, we want to highlight the packages \href{https://gitlab.inria.fr/AlgebraicGeometricModeling/AlgebraicSolvers.jl/-/tree/master/}{\toolbox{AlgebraicSolvers.jl}} by~\cite{telen2018solving} and \href{https://github.com/MBender/JuliaEigenvalueSolver}{\toolbox{JuliaEigenvalueSolver.jl}} by~\cite{bender2022eigenvalue}, which tackle the polynomial systems by using the Macaulay (or a related) matrix to set up the multiplication matrices.

       
    \subsection{Rectangular multiparameter eigenvalue-finding}
        \label{sec:eigenvalueproblem}

        Another problem that can be deduced from~\eqref{eq:unifyingproblem} is the rectangular multiparameter eigenvalue problem. 
        By setting $s = 1$, there is only one matrix equation and $k$ has to be larger or equal to $l + n - 1$ in order to have a well-determined problem.

        \subsubsection{Rectangular multiparameter eigenvalue-finding problem}

            The rectangular multiparameter eigenvalue problem is a generalization of the one-parameter polynomial eigenvalue problem with multiple spectral parameters.
            It is given by a multivariate polynomial matrix.
            The solutions of the rectangular multiparameter eigenvalue problem, or the eigenvalues of the polynomial matrix, are the points $\vc{x}^\ast \in \mathbb{C}^m$ for which the rank of the polynomial matrix drops below the normal rank,
            \begin{equation}
                \rank \mleft( \mpoly(\vc{x}^\ast) \mright) < \nrank \mleft( \mpoly(\vc{x}) \mright).
            \end{equation}
            Every eigenvalue $\vc{x}^\ast$ has at least one eigenvector $\vc{y}^\ast \in \mathbb{C}^{l \times 1}_0$ associated to it.
            The rectangular multiparameter eigenvalue problem is given in the following definition.

            \begin{definition}
                \label{def:multiparametereigenvalueproblem}
                Given coefficient matrices $\mt{A}^{(\vc{i})} \in \mathbb{C}^{k \times l}$ (with $k \geq l + m - 1$) that lead to a full normal rank matrix polynomial $\mpoly(\vc{x})$, the \emph{rectangular multiparameter eigenvalue-finding problem} consists in finding all $m$-tuples $\vc{x}^\ast \in \mathbb{C}^m$ and corresponding vectors $\vc{y}^\ast \in \mathbb{C}^{l \times 1}_0$, so that $\mpoly(\vc{x}^\ast) \vc{y}^\ast = \vc{0}$, with $\left\Vert \vc{y}^\ast \right\Vert = 1$ as normalization constraint.
            \end{definition}

            \begin{example}
                Let the coefficient matrices $\mt{A}^{(\vc{i})}$ be $k \times l$ rectangular matrices.
                A rectangular, quadratic two-parameter eigenvalue problem is then given by
                \begin{equation}
                    \mcl{P} (x_1, x_2) \, \vc{y} = \left( \mt{A}^{(20)} x_1^2 + \mt{A}^{(11)} x_1 x_2 + \mt{A}^{(02)} x_2^2 + \mt{A}^{(10)} x_1 + \mt{A}^{(01)} x_2 + \mt{A}^{(00)} \right) \vc{y} = \vc{0},
                \end{equation}
                where the solutions $(x_1^\ast, x_2^\ast)$ are the eigenvalues and $\vc{y}^\ast \in \mathbb{C}_0^{l \times 1}$ with $\left\Vert \vc{y} \right\Vert = 1$ are the associated eigenvectors.
            \end{example}
                
            Rectangular multiparameter eigenvalue problems appear when solving certain differential equations, such as the higher-order Heine--Stieltjes problem~\cite{shapiro2010algebro} or when tackling the minimum rank problem in cryptoanalysis~\cite{faugere2008cryptanalysis}. 
            Recently, these problems have also emerged as a globally optimal approach to identify the parameters of misfit-latency models~\cite{demoor2019least,demoor2020least,vermeersch2019globally} and to reduce the order of single-input, single-output models given by their transfer function~\cite{agudelo2021globally,lagauw2023globally}.

        \subsubsection{Existing eigenvalue-finding software}

            \macaulaylab is one of the first toolboxes focussing on rectangular multiparameter eigenvalue problems. 
            Very recently,~\cite{hochstenbach2024numerical} have shown that it is possible to translate (linear) rectangular problems into a square formulation. 
            This transformation, together with the available methods for solving square multiparameter eigenvalue problems, is added to \toolbox{MultiParEig} in one of its last updates\footnote{Methods to solve linear rectangular multiparameter eigenvalue problems were added to \toolbox{MultiParEig} with the release of version 2.7.0.0 (December 6, 2022).}, adding a second toolbox to the list of software packages that can solve rectangular multiparameter eigenvalue problems.
            This toolbox by Plestenjak~\cite{plestenjak2023multipareig} features a wide array of algorithms (both direct and iterative) and can deal with singular and non-singular problems.
            When the problem is not linear, however, a linearization step is required, while \macaulaylab deals with polynomial problems directly.
            For problems with a low number of spectral parameters and large coefficient matrices, \toolbox{MultiParEig} is currently the fastest available approach.
            \macaulaylab closes this gap when the number of spectral parameters increases.

\section{Common (block) Macaulay matrix approach}
    \label{sec:blockmacaulaymatrix}

    We explain in this section how the above-described key problems can be addressed via one common approach. 
    Both key problems are transformed via the (block) Macaulay matrix into a multidimensional realization problem, resulting in a joint generalized eigenvalue problem that yields the solutions (\cref{sec:transformation}). 
    This multidimensional realization problem can be constructed from the (right) null space or the column space of the (block) Macaulay matrix (\cref{sec:subspace}).
    In all cases, particular attention must be paid to multiple solutions and solutions at infinity (\cref{sec:multiplicities&infinity}).

    \subsection{Transformation into a joint generalized eigenvalue problem}
        \label{sec:transformation}

\begin{figure}
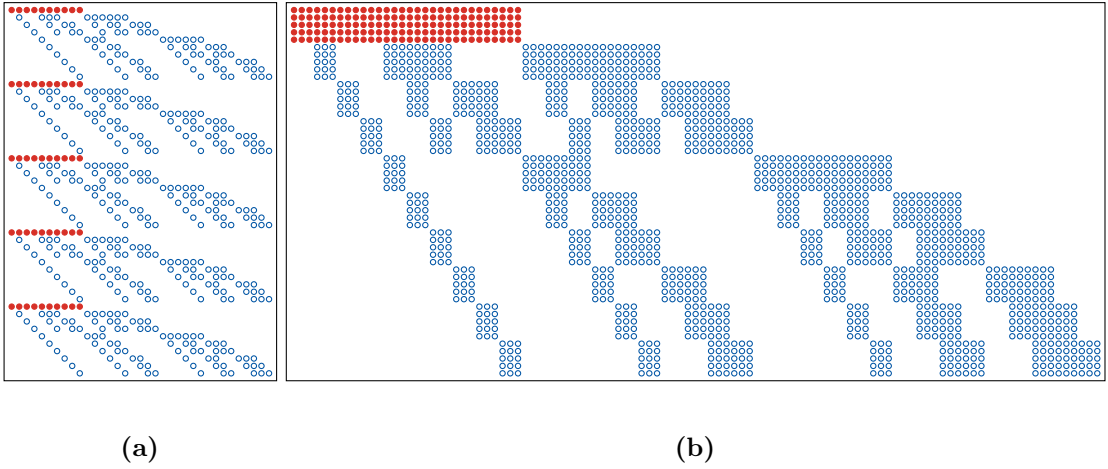

            \centering
            \begin{subfigure}[t]{0.246\textwidth}
                \centering
                \input{figures/macaulayspy.tikz}
                \caption{}
                \label{fig:macaulaymatrix:macaulayspy}
            \end{subfigure}
            \begin{subfigure}[t]{0.739\textwidth}
                \centering
                \input{figures/blockmacaulayspy.tikz}
                \caption{}
                \label{fig:macaulaymatrix:blockmacaulayspy}
            \end{subfigure}
            \ifarxiv \else
                \Description{Visualization of the (block) Macaulay matrix for a multivariate polynomial system and a rectangular multiparameter eigenvalue problem. The (block) Macaulay matrix is constructed from the coefficients of the polynomials or the coefficient matrices of the rectangular multiparameter eigenvalue problem. The original coefficients are indicated by solid red dots, while the open blue dots are the result of the multiplication of the problem with the different monomials.}
            \fi
            \caption{Visualization of the (block) Macaulay matrix for a multivariate polynomial system (\cref{fig:macaulaymatrix:macaulayspy}) and a rectangular multiparameter eigenvalue problem (\cref{fig:macaulaymatrix:blockmacaulayspy}). The (block) Macaulay matrix is constructed from the coefficients of the polynomials or the coefficient matrices of the rectangular multiparameter eigenvalue problem. The original coefficients are indicated by solid red dots \displaylabel{plot:reddot}, while the open blue dots \displaylabel{plot:bluedot} are the result of the multiplication of the problem with the different monomials.}
            \label{fig:macaulaymatrix}
        \end{figure}

        Independently of the fact whether the matrix equations $\mpoly[j] (\vc{x}) \, \vc{y} = \vc{0}$ are multivariate polynomials or rectangular matrix polynomials, the (block) Macaulay matrix approach starts with multiplying the matrix equations by different monomials $\vc{x}^{\vc{i}}$ up to a certain degree $d - d_j$, resulting in a set of matrix equations $\vc{x}^{\vc{i}} \mpoly[j] (\vc{x}) \, \vc{y} = \vc{0}$ of total degree $d$ or smaller.
        The coefficients or coefficient matrices of these matrix equations are arranged in a structured matrix.
        For example, 
        \begin{equation}
            \label{eq:macaulaymatrix}
            \begin{bNiceMatrix}[first-row, code-for-first-row = \color{gray}\scriptstyle, first-col, code-for-first-col = \color{gray}\scriptstyle]
                & \substack{\vc{y} \\ \downarrow} & \substack{x_1 \vc{y} \\ \downarrow} & \substack{x_2 \vc{y} \\ \downarrow} & \substack{x_1^2 \vc{y} \\ \downarrow} & \substack{x_1 x_2 \vc{y} \\ \downarrow} & \substack{x_2^2 \vc{y} \\ \downarrow} & \substack{x_1^3 \vc{y} \\ \downarrow} & \substack{x_1^2 x_2 \vc{y} \\ \downarrow} & \substack{x_1 x_2^2 \vc{y} \\ \downarrow} & \substack{x_2^3 \vc{y} \\ \downarrow} \\[0.2ex]
                \mpoly[1] \, \vc{y} \rightarrow & \mt{A}^{(00)}_{1} & \mt{A}^{(10)}_{1} & \mt{A}^{(01)}_{1} & \mt{A}^{(20)}_{1} & \mt{A}^{(11)}_{1} & \mt{A}^{(02)}_{1} & \mt{0} & \mt{0} & \mt{0} & \mt{0} \\
                x_1 \mpoly[1] \, \vc{y} \rightarrow & \mt{0} & \mt{A}^{(00)}_{1} & \mt{0} & \mt{A}^{(10)}_{1} & \mt{A}^{(01)}_{1} & \mt{0} & \mt{A}^{(20)}_{1} & \mt{A}^{(11)}_{1} & \mt{A}^{(02)}_{1} & \mt{0} \\
                x_2 \mpoly[1] \, \vc{y} \rightarrow & \mt{0} & \mt{0} & \mt{A}^{(00)}_{1} & \mt{0} & \mt{A}^{(10)}_{1} & \mt{A}^{(01)}_{1} & \mt{0} & \mt{A}^{(20)}_{1} & \mt{A}^{(11)}_{1} & \mt{A}^{(02)}_{1} \\
                \mpoly[2] \, \vc{y} \rightarrow & \mt{A}^{(00)}_{2} & \mt{A}^{(10)}_{2} & \mt{A}^{(01)}_{2} & \mt{A}^{(20)}_{2} & \mt{A}^{(11)}_{2} & \mt{A}^{(02)}_{2} & \mt{0} & \mt{0} & \mt{0} & \mt{0} \\
                x_1 \mpoly[2] \, \vc{y} \rightarrow & \mt{0} & \mt{A}^{(00)}_{2} & \mt{0} & \mt{A}^{(10)}_{2} & \mt{A}^{(01)}_{2} & \mt{0} & \mt{A}^{(20)}_{2} & \mt{A}^{(11)}_{2} & \mt{A}^{(02)}_{2} & \mt{0} \\
                x_2 \mpoly[2] \, \vc{y} \rightarrow & \mt{0} & \mt{0} & \mt{A}^{(00)}_{2} & \mt{0} & \mt{A}^{(10)}_{2} & \mt{A}^{(01)}_{2} & \mt{0} & \mt{A}^{(20)}_{2} & \mt{A}^{(11)}_{2} & \mt{A}^{(02)}_{2}
            \end{bNiceMatrix}
        \end{equation}
        is a (block) Macaulay matrix constructed from two matrix equations, $\mpoly[1](x_1, x_2)$ and $\mpoly[2](x_1, x_2)$, and multiplied by monomials $x_1^{i_1} x_2^{i_2}$, where $d = 3$.
        The block rows correspond to the different matrix equations, while the block columns are labeled by the different monomials up to degree $d$. 
        The specific structure of the (block) Macaulay matrix depends on the polynomial basis and monomial order that is used. 
        \Cref{fig:macaulaymatrix} shows a (block) Macaulay matrix for a multivariate polynomial system and a rectangular multiparameter eigenvalue problem.

        The (block) Macaulay matrices form a family of structured matrices $\mt{M}(d) \in \mathbb{C}^{p \times q}$ parametrized by its degree $d$.
        If the degree $d$ of the (block) Macaulay matrix is large enough, then its nullity $n$ corresponds to the number of solutions of~\eqref{eq:unifyingproblem} for zero-dimensional solution sets~\cite{vermeersch2021column,vermeersch2022two,vermeersch2023block}.
        In that case, the null space of $\mt{M}$ has a shift-invariant structure that can be exploited to retrieve these solutions.
        One possible basis matrix $\mt{V} \in \mathbb{C}^{q \times n}$ for this null space consists of one (block) multivariate Vandermonde vector $\vc{v}$ for every solution, for example,
        \begin{equation}
            \vc{v} = 
            \begin{bmatrix}
                \vc{y}^\tr & x_1 \vc{y}^\tr & x_2 \vc{y}^\tr & x_1^2 \vc{y}^\tr & x_1 x_2 \vc{y}^\tr & x_2^2 \vc{y}^\tr & x_1^3 \vc{y}^\tr & x_1^2 x_2 \vc{y}^\tr & x_1 x_2^2 \vc{y}^\tr & x_2^3 \vc{y}^\tr
            \end{bmatrix}^\tr
        \end{equation}
        evaluated in each of the solutions $(\vc{x}^\ast, \vc{y}^\ast)$.
        Shift-invariance means that we can select rows of a basis matrix of the null space, multiply them by a polynomial in $\vc{x}$, and find the result again as a combination of rows in that same basis matrix:
        \begin{equation}
            \begin{bmatrix}
                \vc{y} \\
                x_1 \vc{y} \\
                x_2 \vc{y}
            \end{bmatrix} 
            \rightarrow 
            \begin{bmatrix}
                x_1 \vc{y} \\
                x_1^2 \vc{y} \\
                x_1 x_2 \vc{y} 
            \end{bmatrix}
            + 2
            \begin{bmatrix}
                x_2 \vc{y} \\
                x_1 x_2 \vc{y} \\
                x_2^2 \vc{y} 
            \end{bmatrix},
        \end{equation}
        for shifting three rows by the polynomial $x_1 + 2 x_2$.
        The shift operation corresponds to
        \begin{equation}
            \mt{S}_0 \mt{V} \mt{D}_i = \mt{S}_i \mt{V},
        \end{equation}
        where the matrices $\mt{S}_0 \in \mathbb{C}^{n \times q}$ and $\mt{S}_i \in \mathbb{C}^{n \times q}$ select rows from the basis matrix $\mt{V}$, before and after multiplication with the polynomial evaluated in $\mt{D}_i \in \mathbb{C}^{n \times n}$.
        Recall that the basis matrix $\mt{V}$ consists of (block) multivariate Vandermonde vectors constructed from the solutions of the problem and the linearly independent rows that we select are related to the solutions of the problem.
        In practice, we do not know the solutions and obtain a basis matrix $\mt{Z}$ for the null space of the (block) Macaulay matrix via numerical linear algebra. 
        Both basis matrices are related by a nonsingular transformation matrix $\mt{T} \in \mathbb{C}^{n \times n}$, resulting in a different, solvable relation
        \begin{equation}
            \label{eq:geps}
            \left(\mt{S}_0 \mt{Z}\right) \mt{T} \mt{D}_i = \left(\mt{S}_i \mt{Z}\right) \mt{T}.
        \end{equation} 
        These matrix pairs, $\left(\mt{S}_i \mt{Z}, \mt{S}_0 \mt{Z}\right)$, which we obtain from shift operations with different monomials $x_i, i = 1, \ldots, m$ are $m$ generalized multiplication matrices, the $n$ tuples of eigenvalues of which constitute the solutions of the problem in~\eqref{eq:unifyingproblem}. 

        The selection of the rows in the basis matrix $\mt{Z}$ via $\mt{S}_0$ is so that the monomials that correspond to the these rows form a basis for the coordinate ring~\cite{bender2022eigenvalue}.
        Gr\"{o}bner or border basis methods are used to find these monomials in computational algebraic geometry~\cite{stetter2004numerical}.
        Here, we search for the linearly independent rows (checked from top to bottom) in the basis matrix $\mt{Z}$ of the null space and select them via $\mt{S}_0$, which is similar to finding the standard monomials associated to the affine solutions of the problem.
        The selection of standard monomials makes sure that~\eqref{eq:geps} is a consistent generalized eigenvalue problem.

    \subsection{Null spaces versus column space as solution subspace}
        \label{sec:subspace}

        The above-described procedure is called the \emph{null space based approach}. 
        A similar procedure exists that uses the column space of the (block) Macaulay matrix to construct the generalized multiplication matrices.
        It starts from the linear algebra observation that every linearly independent row in the null space corresponds with a linearly dependent column in the column space~\cite{vermeersch2021column}. 
        Such linearly independent row of the null space and linearly dependent column of the column space correspond to the same solution of the problem, and the generalized multiplication matrices need these rows/columns in their construction.

        Instead of constructing the generalized eigenvalue problems from the linearly independent row of the null space, the \emph{column space based approach} uses the linearly dependent columns of the (block) Macaulay matrix, avoiding the construction of a basis of the null space~\cite{vermeersch2021column,vermeersch2022two}.
        The generalized multiplication matrices are constructed as 
        \begin{equation}
            \label{eq:columnspace}
            \left(\mt{R}_{33} \right) \mt{T} \mt{D}_i = \left(- \mt{R}_ {34}\right) \mt{T}.
        \end{equation}
        where the matrices $\mt{R}_{33}$ and $\mt{R}_{34}$ are blocks of the upper triangular matrix obtained after performing a backward QR decomposition of the (block) Macaulay matrix $\mt{M}$, in which the columns are re-ordered such that the linearly dependent columns that correspond to the affine solutions (cf., the rows selected via $\mt{S}_0$) and columns after the shift operation (cf., the rows selected via $\mt{S}_i$) are positioned at the left side of the matrix.

    \subsection{Influence of multiplicities and solutions at infinity}
        \label{sec:multiplicities&infinity}

        Constructing the generalized eigenvalue problems in~\eqref{eq:geps} is only possible for problems with affine and simple solutions. 
        Solutions at infinity need to be removed from the solution space before setting up the shift problems, while multiple solutions have a negative impact on the accuracy.

        \paragraph*{Deflating solutions at infinity}

\begin{figure}
                \centering
                \input{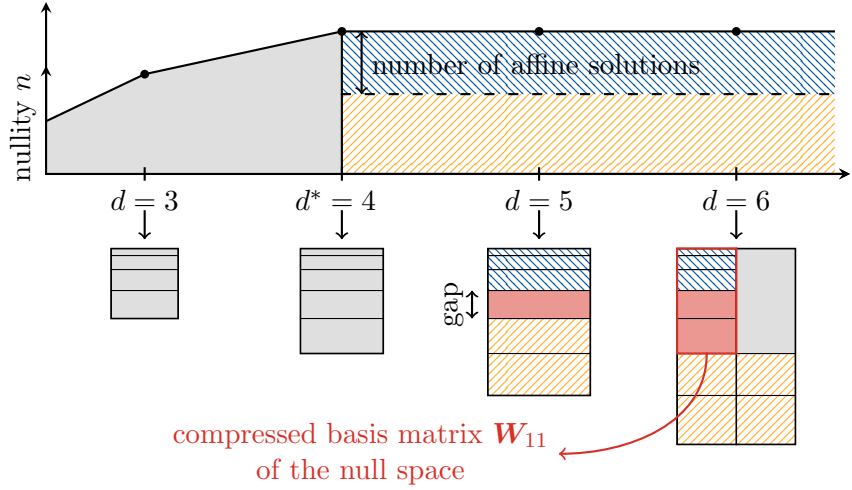}
                \ifarxiv \else
                    \Description{Structure of the null space}
                \fi
                \caption{Visualization of the structure of a basis matrix $\mt{Z}$ of the null space of the (block) Macaulay matrix when increasing the degree $d$. More (block) rows are added in every degree, in accordance with the growing (block) Macaulay matrix. From a certain degree on, $d^\ast$ in this example, the nullity $n$ corresponds to the number of solutions. By checking the rank structure, it is possible to split the affine solutions and solutions at infinity. There is a gap zone that separates both types of solutions. A column compression can be used to deflate the solutions at infinity.}
                \label{fig:nullspace}
            \end{figure}

            The solutions at infinity can be deflated via a column compression of the null space, in which the basis matrix $\mt{Z}$ is compressed to a smaller matrix $\mt{W}_{11} \in \mathbb{C}^{n \times n}$~\cite{vermeersch2022two,vermeersch2023block}.
            The column compression computes a basis for the column space of $\mt{Z}$ that corresponds to the affine solutions~\cite{dreesen2013back}.
            Setting up the multiplication maps from the compressed null space $\mt{W}_{11}$,
            \begin{equation}
                \label{eq:nullspace}    
                \left(\mt{S}_0 \mt{W}_{11}\right) \mt{T} \mt{D}_i = \left(\mt{S}_i \mt{W}_{11}\right) \mt{T}.
            \end{equation}
            leads to generalized eigenvalue problems that only contains the affine solutions.
            This requires rank checks to determine the rank structure (i.e., checking how many of the solutions in the null space are affine and where they are situated). 
            A typical rank structure of the null space is visualized in \cref{fig:nullspace}, showing how the solutions at infinity can be identified via a gap zone and removed.
            For the column space based approach, the solutions at infinity are removed implicitly by the backward QR decomposition that sets up the multiplication maps in~\eqref{eq:columnspace}.

        \paragraph*{Positive-dimensional solution set}
            An important condition for obtaining correct solutions is that the nullity of the (block) Macaulay matrix is equal to the number of solutions. 
            This means that the number of solutions must be finite, i.e., the solution space must be zero-dimensional, otherwise the nullity  can not reach this number.
            Solvers that use Macaulay-like matrices have therefore the important restriction that they can only solve problems with a zero-dimensional solution space.
            There is an exception to this rule for \macaulaylab: If the positive-dimensional part of the solution space is entirely situated at infinity, then the null space of the (block) Macaulay matrix after column compression is still zero-dimensional and the algorithm can be used to find the isolated affine solutions (with a careful implementation).

        \paragraph*{Clustering of multiple solutions}

            Furthermore, the obtained affine solutions, being it from the null space or column space, can have multiplicities. 
            These multiplicities cause the accuracy of the algorithm to drop.
            The (block) Macaulay matrix algorithm in \macaulaylab uses a clustering step to identify and refine these multiple solutions, improving their accuracy~\cite{corless1997reordered,vermeersch2023block}.

\section{Structure of the software package}
    \label{sec:structure}

    \begin{figure}
        \begin{tikzpicture}[remember picture]
            \node (algorithm) at (0,0) {
                \begin{minipage}{\textwidth-8pt}
                        \begin{algorithmic}[1]
                            \ifarxiv \else
                                \sffamily
                            \fi
                            \small
                            \Require{coefficients or coefficient matrices of the problem (+ options for the algorithm)}~\tikzmark{input}
                            \While{the solution subspace is not yet converged}
                                \State Increase the degree: $d \leftarrow d + 1$
                                \State Enlarge the solution subspace by constructing $\mt{M}$ or $\mt{Z}$ for degree $d$~\tikzmark{step1}
                                \State Check the rank structure of the solution subspace $\mt{M}$ or $\mt{Z}$~\tikzmark{step2}
                            \EndWhile
                            \If{the solution subspace is the null space} \Comment{null space based approach}
                                \State Perform the column compression to construct $\mt{W}_{11}$~\tikzmark{step3}
                                \State Construct multiplication maps~\eqref{eq:nullspace} from the compressed null space~\tikzmark{step4top}
                            \Else \Comment{column space based approach}
                                \State Construct multiplication maps~\eqref{eq:columnspace} from the column space~\tikzmark{step4}
                            \EndIf
                            \State Compute the eigenvalues $\vc{x}^\ast$ of the multiplication maps~\tikzmark{step4bottom}
                            \State Cluster the affine solutions (optional)~\tikzmark{step5}
                            \State Evaluate the residual errors and compute  vectors $\vc{y}^\ast$ (optional)~\tikzmark{step6}
                            \Ensure{affine solutions and residual errors (+ obtained information on the problem)}~\tikzmark{output}
                        \end{algorithmic}
                \end{minipage}
            };
            
            \draw[line width = \heavyrulewidth] (algorithm.south west) -- (algorithm.south east);
            \draw[line width = \heavyrulewidth] (algorithm.north west) -- (algorithm.north east);
            
            \coordinate (node1) at ($(pic cs:step1)+ (0,0.5ex)$);
            \coordinate (node2) at ($(pic cs:step2)+ (0,0.5ex)$);
            \coordinate (node3) at ($(pic cs:step3)+ (0,0.5ex)$);
            \coordinate (node4top) at ($(pic cs:step4top)+ (0,1.75ex)$);
            \coordinate (node4) at ($(pic cs:step4)+ (0,0.5ex)$);
            \coordinate (node4bottom) at ($(pic cs:step4bottom)+ (0,-0.75ex)$);
            \coordinate (node5) at ($(pic cs:step5)+ (0,0.5ex)$);
            \coordinate (node6) at ($(pic cs:step6)+ (0,0.5ex)$);
            \node[font = {\sffamily\small\bfseries}, anchor = east] (text1) at (node1 -| algorithm.east) {Step~\ctext{1}};
            \node[font = {\sffamily\small\bfseries}, anchor = east] (text2) at (node2 -| algorithm.east) {Step~\ctext{2}};
            \node[font = {\sffamily\small\bfseries}, anchor = east] at (node3 -| algorithm.east) {Step~\ctext{3}};
            \draw[font = {\sffamily\small\bfseries}, thick, decorate, decoration = {brace, amplitude = 3pt}] ($(node4top -| algorithm.east) - (1.5cm,0)$) -- ($(node4bottom -| algorithm.east) - (1.5cm,0)$);
            \node[font = {\sffamily\small\bfseries}, anchor = east] at (node4 -| algorithm.east) {Step~\ctext{4}};
            \node[font = {\sffamily\small\bfseries}, anchor = east] at (node5 -| algorithm.east) {Step~\ctext{5}};
            \node[font = {\sffamily\small\bfseries}, anchor = east] at (node6 -| algorithm.east) {Step~\ctext{6}};
        \end{tikzpicture}
        \ifarxiv \else
            \Description{Same as below?}
        \fi
        \caption{Pseudocode for the (block) Macaulay matrix approach, as described in \cref{sec:blockmacaulaymatrix}. Six coherent steps can be deduced from the pseudocode, which are explained and linked to the implementation of \macaulaylab in \cref{sec:structure}.}
        \label{fig:pseudocode}
    \end{figure}

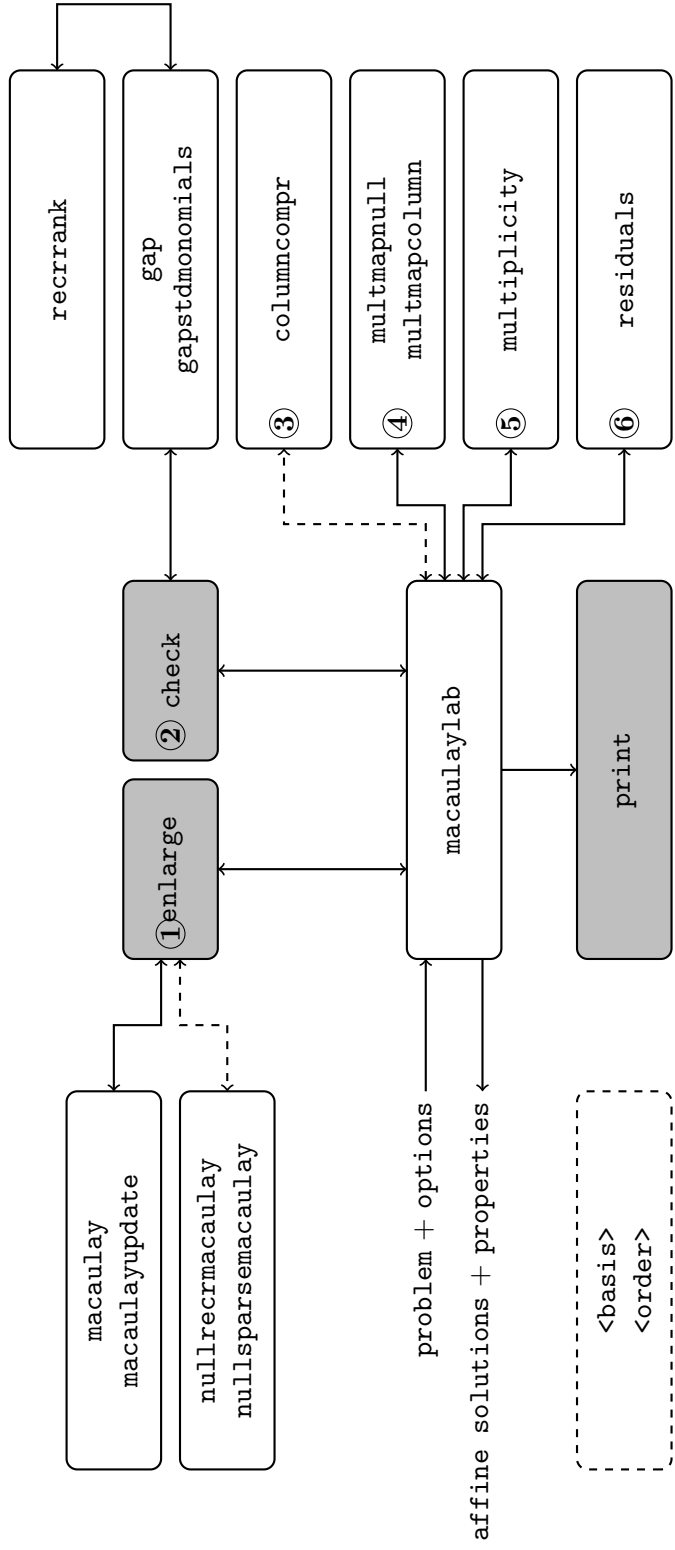
\begin{sidewaysfigure}
        \centering
        \hspace*{\fill}
        \begin{tikzpicture}
    \node[anchor = east] at (-1.4\cu,-1.0\cu) {\texttt{problem} + \texttt{options}};
    \draw[thick, ->] (-1.4\cu,-1.0\cu) -- (0,-1.0\cu);
    \node[anchor = east] at (-1.4\cu,-1.6\cu) {\texttt{affine solutions} + \texttt{properties}};
    \draw[thick, ->] (0,-1.6\cu) -- (-1.4\cu,-1.6\cu);

    \draw[thick, rounded corners] (0,-1.8\cu) rectangle (4\cu,-0.8\cu);
    \node at (2\cu,-1.3\cu) {\texttt{macaulaylab}};

    \draw[thick, rounded corners, fill = gray!50] (0,-3.6\cu) rectangle (4\cu,-2.6\cu);
    \node at (2\cu,-3.1\cu) {\texttt{print}};

    \draw[thick, rounded corners] (5.4\cu,2.4\cu) rectangle (9.4\cu,3.4\cu);
    \node at (7.4\cu,2.9\cu) {\texttt{recrrank}};

    \draw[thick, rounded corners] (5.4\cu,1.2\cu) rectangle (9.4\cu,2.2\cu);
    \node[align = center] at (7.4\cu,1.7\cu) {\texttt{gap} \\ \texttt{gapstdmonomials}};

    \draw[thick, rounded corners] (5.4\cu,0) rectangle (9.4\cu,\cu);
    \node at (7.4\cu,0.5\cu) {\texttt{columncompr}};
    \node[font = {\bfseries}, anchor = west] at (5.4\cu,0.5\cu) {\ctext{3}};

    \draw[thick, rounded corners] (5.4\cu,-1.2\cu) rectangle (9.4\cu,-0.2\cu);
    \node[align = center] at (7.4\cu,-0.7\cu) {\texttt{multmapnull} \\ \texttt{multmapcolumn}};
    \node[font = {\bfseries}, anchor = west] at (5.4\cu,-0.7\cu) {\ctext{4}};

    \draw[thick, rounded corners] (5.4\cu,-2.4\cu) rectangle (9.4\cu,-1.4\cu);
    \node at (7.4\cu,-1.9\cu) {\texttt{multiplicity}};
    \node[font = {\bfseries}, anchor = west] at (5.4\cu,-1.9\cu) {\ctext{5}};

    \draw[thick, rounded corners] (5.4\cu,-3.6\cu) rectangle (9.4\cu,-2.6\cu);
    \node at (7.4\cu,-3.1\cu) {\texttt{residuals}};
    \node[font = \bfseries, anchor = west] at (5.4\cu,-3.1\cu) {\ctext{6}};

    \draw[thick, rounded corners, fill = gray!50] (2.1\cu,1.2\cu) rectangle (4\cu,2.2\cu);
    \node at (3.05\cu,1.7\cu) {\texttt{check}};
    \node[font = {\bfseries}, anchor = west] at (2.1\cu,1.7\cu) {\ctext{2}};

    \draw[thick, rounded corners, fill = gray!50] (0,1.2\cu) rectangle (1.9\cu,2.2\cu);
    \node at (0.95\cu,1.7\cu) {\texttt{enlarge}};
    \node[font = {\bfseries}, anchor = west] at (0,1.7\cu) {\ctext{1}};

    \draw[thick, rounded corners] (-5.4\cu,1.8\cu) rectangle (-1.4\cu,2.8\cu);
    \node[align = center] at (-3.4\cu,2.3\cu) {\texttt{macaulay} \\ \texttt{macaulayupdate}};

    \draw[thick, rounded corners] (-5.4\cu,0.6\cu) rectangle (-1.4\cu,1.6\cu);
    \node[align = center] at (-3.4\cu,1.1\cu) {\texttt{nullrecrmacaulay} \\ \texttt{nullsparsemacaulay}};

    \draw[thick, rounded corners, dashed] (-5.4\cu,-3.6\cu) rectangle (-1.4\cu,-2.6\cu);
    \node[align = center] at (-3.4\cu,-3.1\cu) {\texttt{<basis>} \\ \texttt{<order>}};

    \draw[thick, <->] (4\cu,1.7\cu) -- (5.4\cu,1.7\cu);
    \draw[thick, <->] (3.05\cu,1.2\cu) -- (3.05\cu,-0.8\cu);
    \draw[thick, <->] (0.95\cu,1.2\cu) -- (0.95\cu,-0.8\cu);
    \draw[thick, ->] (2\cu,-1.8\cu) -- (2\cu,-2.6\cu);
    \draw[thick, <->] (0,1.8\cu) -- (-0.7\cu,1.8\cu) -- (-0.7\cu,2.3\cu) -- (-1.4\cu,2.3\cu);
    \draw[thick, <->, dashed] (0,1.6\cu) -- (-0.7\cu,1.6\cu) -- (-0.7\cu,1.1\cu) -- (-1.4\cu,1.1\cu);

    \draw[thick, <->, dashed] (4\cu,-1.0\cu) -- (4.6\cu,-1.0\cu) -- (4.6\cu,0.5\cu) -- (5.4\cu,0.5\cu);

    \draw[thick, <->] (4\cu,-1.2\cu) -- (4.8\cu,-1.2\cu) -- (4.8\cu,-0.7\cu) -- (5.4\cu,-0.7\cu);
    \draw[thick, <->] (4\cu,-1.4\cu) -- (4.8\cu,-1.4\cu) -- (4.8\cu,-1.9\cu) -- (5.4\cu,-1.9\cu);
    \draw[thick, <->] (4\cu,-1.6\cu) -- (4.6\cu,-1.6\cu) -- (4.6\cu,-3.1\cu) -- (5.4\cu,-3.1\cu);
    \draw[thick, <->] (9.4\cu,1.7\cu) -- (10.1\cu,1.7\cu) -- (10.1\cu,2.9\cu) -- (9.4\cu,2.9\cu);
\end{tikzpicture}
        \hspace*{\fill}
        \caption{Diagram of the main functions (rectangles) implemented in \macaulaylab and their dependencies (arrows). The gray rectangles indicate local functions from \function{macaulaylab} and the dashed arrows denote dependencies that are only used in the null space based approach. Since the solution approach is similar for both problem types, the same six steps are used and most functions are capable of dealing directly with both problem types (scalars coefficients are replaced by coefficient matrices, and vice versa). Note that most functions depend on the chosen polynomial basis (implemented in \function{<basis>}) and monomial order (implemented in \function{<order>}).}
        \label{fig:diagram}
    \end{sidewaysfigure}

    The different steps of the (block) Macaulay matrix approach are outlined in the pseudocode shown in \cref{fig:pseudocode}, and they are reflected in the implementation of the toolbox.
    \macaulaylab solves both key problems in six steps: 
    \begin{itemize}
        \item Step~\ctext{1}: enlarge the solution subspace by increasing the degree,
        \item Step~\ctext{2}: check whether the solution subspace can accommodate the shift,
        \item Step~\ctext{3}: remove the solutions at infinity,
        \item Step~\ctext{4}: exploit the shift-invariance of the solution subspace,
        \item Step~\ctext{5}: cluster the affine solutions to obtain a better accuracy (optional), and
        \item Step~\ctext{6}: compute the residual errors of the obtained affine solutions.
    \end{itemize}
    \Cref{fig:diagram} visualizes the main functions in \macaulaylab that enable these different steps and shows how they interact with each other.
    (Notice that \function{enlarge} and \function{check} are two local functions inside the solver \function{macaulaylab}.)
    While most of the time, the same functions can deal with both problem types, this requires a careful implementation that takes into account the subtle differences between them.
    The toolbox is built in a modular fashion, which allows easy expansions and improvements in future updates. 
    Note that many of the functions depend on the chosen polynomial basis (implemented in \function{<basis>}) and monomial order (implemented in \function{<position>}), more information about this feature is provided in~\cref{sec:basis&order}.

    In the remainder of this section, we dive deeper into the six different steps required to solve both key problems via the (block) Macaulay matrix approach.
    We deal with one step at a time and highlight the different implementation decisions.
    It is important to note that implementation decisions always represent making a substantiated choice in the trade-off between computational efficiency and numerical robustness.

    \step{Define the problem in the correct format}
        \label{step:initialstep}

\begin{figure}[t]
            \centering
            \begin{tikzpicture}
    \draw[thick, pattern = north west lines, pattern color=gray, rounded corners] (-0.2\cu,-2.2\cu) rectangle (9.2\cu,1.2\cu);
    \node[text width = 4\cu, align = center, anchor = south] at (2\cu,1.5\cu) {multivariate polynomial system with equations \texttt{eqs}};
    \node[text width = 4\cu, align = center, anchor = south] at (7\cu,1.5\cu) {\texttt{m}-parameter eigenvalue problem with coefficient matrices \texttt{mat} and support \texttt{supp}};

    \draw[->, thick] (2\cu,1.5\cu) -- (2\cu,\cu);
    \draw[->, thick] (7\cu,1.5\cu) -- (7\cu,\cu);

    \draw[thick, fill = white, rounded corners] (0,0) rectangle (4\cu,\cu);
    \draw[thick, fill = white, rounded corners] (5\cu,0) rectangle (9\cu,\cu);
    \node at (2\cu,0.5\cu) {\texttt{systemstruct(eqs)}};
    \node[anchor = south, yshift = -2pt] at (2\cu,0) {inherits from \texttt{problemstruct}};
    \node[anchor = south, yshift = -2pt] at (7\cu,0) {inherits from \texttt{problemstruct}};
    \node at (7\cu,0.5\cu) {\texttt{mepstruct(mat,supp)}};

    \draw[thick, fill = white, rounded corners] (2.5\cu,-2\cu) rectangle (6.5\cu,-1\cu);
    \node at (4.5\cu,-1.5\cu) {\texttt{problemstruct(coef,supp)}};

    \draw[->, thick, dashed] (2\cu,0) -- (2\cu,-0.5\cu) -- (4.4\cu,-0.5\cu) -- (4.4\cu,-1\cu);

    \node[anchor = south] at (3.2\cu,-0.5\cu) {\texttt{system}};
    \node[anchor = south] at (5.8\cu,-0.5\cu) {\texttt{mep}};

    \draw[->, thick, dashed] (7\cu,0) -- (7\cu,-0.5\cu) -- (4.6\cu,-0.5\cu) -- (4.6\cu,-1\cu);

    \node[anchor = north, align = left] at (7.5\cu,-2.5\cu) {internal representation};
    \draw[->, thick] (7.5\cu,-2.5\cu) -- (7.5\cu,-1.5\cu) -- (6.5\cu,-1.5\cu);

    \draw[->, thick] (4.5\cu,-2\cu) -- (4.5\cu,-2.5\cu);
    \node[anchor = north] at (4.5\cu,-2.5\cu) {\texttt{problem}};
\end{tikzpicture}
            \ifarxiv \else
                \Description{Explain figure!}
            \fi
            \caption{Construction of the data structure to represent a multivariate polynomial system or a rectangular multiparameter eigenvalue problem. Both problem types are internally represented by the same \function{problemstruct}: all necessary information is stored in the cells \function{coef} and \function{supp}. The sub-classes \function{systemstruct} and \function{mepstruct} provide constructors to set-up the problems more easily, but it is also possible to submit the problem directly in its internal representation. It is possible to specify the basis in which the problem is constructed.}
            \label{fig:datastructures}
        \end{figure}
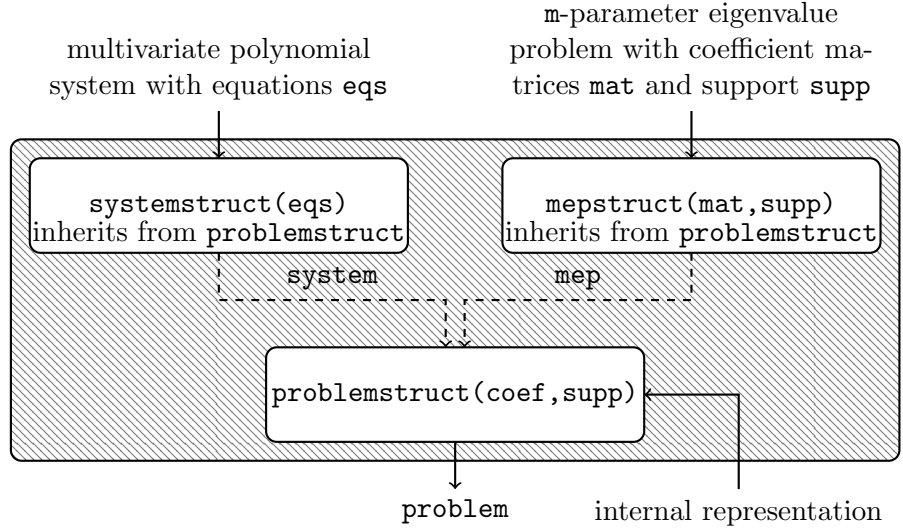

        Both problem types are internally represented by the same class \function{problemstruct}: all necessary information is stored in the cell arrays \function{coef} and \function{supp}, which is a data type in \matlab with indexed data containers, called cells, where each cell can contain any type of data. For \macaulaylab, each cell of \function{coef} and \function{supp} correspond to one polynomial (matrix) equation. A multidimensional array in the corresponding \function{coef} cell contains the coefficients/coefficient matrices, while a two-dimensional array in the \function{supp} cell stores the support of these coefficients/coefficient matrices.  
        Although it is also possible to submit the problem directly in its internal representation, the sub-classes \function{systemstruct} and \function{mepstruct} provide constructors to set-up the specific problems more easily (\cref{fig:datastructures}).
        The uniform internal data structure allows the functions of \macaulaylab to manipulate both types of problems similarly. 


    \step{Enlarge the solution subspace}
        \label{step:enlarge}         

        The first step to build and enlarge the solution subspace until it is large enough, which is checked for every degree in Step~\ctext{2}.
        \macaulaylab supports two solution subspaces: the column space and (right) null space of the (block) Macaulay matrix. 
        The column space can be enlarged iteratively (via \function{macaulay}) or recursively (via \function{macaulayupdate}), in which case the previous (block) Macaulay matrix is used to build the new one.
        Enlarging the (right) null space can be done iteratively (via \function{macaulay} and \function{null}), recursively (via \function{macaulayupdate} and \function{nullrecrmacaulay}), or sparsely (via \function{nullsparsemacaulay}), in which case the (right) null space is determined without constructing the corresponding (block) Macaulay matrix, but by the sparse matrix constructions developed in~\cite{vermeersch2023recursive}.
        It is important to note that, depending on the chosen polynomial basis and monomial order, the resulting matrices can have a very different structure. 
        However, everything is implemented independent of the polynomial basis and monomial order, as we explain in~\cref{sec:basis&order}.


    \step{Check the rank structure}
        \label{step:check}

        Of course, we need to know whether the solution subspace contains a gap zone that separates the affine solutions and solutions at infinity, i.e., whether the degree of the (block) Macaulay subspace is large enough.
        We do this via rank checks of the solution space at a certain degree. 
        There are two approaches implemented to do this: by looking at the increase of the rank for subsequent rows/columns or entire subsequent degree blocks.
        By default, we work block-wise per degree block, since this is numerically more robust and computationally more efficient than row-wise rank checks.

        
        Both the basis matrix of the (right) null space and the flipped transpose of the (block) Macaulay matrix can be interpreted as block row matrices, when we consider the rows/columns degree block-wise. 
        Note that we typically do not immediately check the rank structure for all iterations of the enlargement. 
        For zero-dimensional solution sets, we wait until the nullity is stabilized, and we only start checking the rank structure of the solution space after stabilization. 
        This way, we avoid many superfluous rank checks and create a more efficient solution algorithm.
        When the problem has a positive-dimensional solution space, it is useless to wait for stabilization, since this will never happen (see \cref{sec:multiplicities&infinity}).
        By setting an optional flag (i.e., \function{posdim = true}), the solution algorithm does not wait for the nullity to stabilize, but checks the rank structure of the solution subspace for every degree.

    \step{Perform the column compression}
        \label{step:columncompression}

        In the null space based approach, we need to deflate the solutions at infinity from the (right) null space via a column compression of its basis matrix, which is implemented via a single singular value decomposition.
        Via \function{[U, \codetilde, \codetilde] = svd(Z(1:nrows,:))} and \function{W11 = U(:,1:ma)}, a basis matrix $\mt{W}_{11} \in \mathbb{C}^{n \times ma}$ is constructed that contains the affine solutions, where \function{nrows} is the number of rows until the rows related to solutions at infinity start and \function{ma} is the number of affine solutions.
        Both values are determined in \stepref{2}.

        In the column space based approach, \stepref{3} is not necessary, since the backward QR decompositions in Step~\ctext{4} to set-up the shift problems remove those infinite solutions implicitly. 

    \step{Find eigenvalues of multiplication maps}
        \label{step:multiplicationmaps}

        Given a large-enough solution subspace, the shift problems are set-up as described in~\cref{sec:blockmacaulaymatrix}.
        We consider $m + 1$ shift problems: we shift with a random linear shift polynomial and with the $m$ different solutions components (i.e., the variable coordinates $x_i$, for $i = 1, \ldots, m$).
        Other shift polynomials are possible, but using a random shift polynomial avoids false multiplicities~\cite{corless1997reordered}.
        We use a Schur decomposition to solve the first shift problem and re-use the obtained unitary matrix to obtain the upper triangular matrices of the other $m$ shift problems by pre-multiplication and post-multiplication. 
        This makes sure that the different components of the same solution are at the same position on the diagonal of the upper triangular matrices, which avoids the need for matching the different solution components.
        Note that the $m + 1$ default shift problems only consist of linear shifts. 
        This has a clear advantage: 
        We do only need a gap zone of one degree block. 
        A shift polynomial with a large degree requires a large gap zone, hence more iterations are needed to construct a solution subspace that can accommodate this shift.

        
        When choosing the column space based approach, every shift problem requires one backward QR decomposition to construct the necessary matrices $\mt{R}_{33}$ and $\mt{R}_{34}$ in~\eqref{eq:columnspace}. 
        The backward QR decompositions can be implemented easily (when not considering the structure and sparsity of the (block) Macaulay matrix) as \function{R = qr(fliplr(N)); R = fliplr(R);}. This backward QR decomposition implicitly removes the solutions at infinity. 

        To avoid the sparse multiplication of the row-selection $\mt{S}_0$ and row-combination matrices $\mt{S}_i$ with the subspace matrix when constructing the generalized multiplication matrices, we perform these multiplications indirectly and do not build the row-selection/combination matrices explicitly. 
        It is again important to take into account the correct polynomial basis and monomial order when shifting, otherwise the results are meaningless.

    \step{Cluster the multiple solutions}
        \label{step:cluster}

        An additional, but optional, step in the solution approach is to cluster the obtained affine solutions. 
        We cluster the different solutions based on the evaluations of the random shift polynomial, which are obtained as the eigenvalues of the first shift problem.
        Clustering means that similar evaluations of the shift polynomial are considered to be equal solutions of the problems; hence, we construct clusters of multiple solutions.
        When the shift polynomial is chosen to be random (for other polynomials, we do not perform this clustering step by default), similar evaluations correspond to the same solutions with a probability equal to one.
        Every solution for which the evaluation of the random shift polynomial is grouped in the same cluster, is considered to be the same solution.
        Therefore, we take the geometric mean of the cluster for every component of the solution (i.e., the cluster center).
        The idea for this additional clustering step comes from~\cite{corless1997reordered} in the context of polynomial system solving, but is also very useful when solving rectangular multiparameter eigenvalue problems.


    \step{Compute the residual errors}
        \label{step:residuals}
        
        Finally, we compute the (absolute and relative) residual error for every obtained affine solution:
        \begin{itemize}
            \item For a multivariate polynomial system, we evaluate all the polynomials in the obtained affine solution and sum the absolute values of (residual) values. 
            The relative residual error is obtained by dividing the absolute residual by the sum of the absolute coefficients.
            \item For a rectangular multiparameter eigenvalue problem, we evaluate the matrix polynomial in the obtained affine solutions and compute the associated eigenvector as the right-most right singular vector of the matrix.
            The $2$-norm of the matrix-vector product is the absolute residual, which can again be divided by the coefficients of the problem to obtain the relative residual error.
        \end{itemize}
        In \macaulaylab, we use a clever way to compute both absolute residuals.
        Regardless of the problem type, we compute the smallest singular value of the evaluated problem.
        This value corresponds to the magnitude of the vector of evaluated polynomials when the problem is multivariate polynomial system, while it corresponds to smallest possible residual that we can obtain in the matrix-vector product between the matrix polynomial and its eigenvector.
        Depending on the chosen polynomial basis, a particular evaluation function \function{value} is used to achieve this goal. 


\section{Polynomial basis and monomial order}
    \label{sec:basis&order}

    \macaulaylab is implemented independently from the specific polynomial basis and monomial order. 
    This means that the user can choose the polynomial basis and monomial order that suits the application best, while all the functions keep working out-of-the-box. 
    How is this implemented in the toolbox?
    \begin{itemize}
        \item Every time it is necessary to shift (i.e., multiply) two monomials in the algorithm, the functions call the chosen function \function{<basis>}, which implements the basic shift operation in a particular polynomial basis. By using the correct shift function, the other functions do not have to take into account the specific polynomial basis. For example, in the standard monomial basis, when shifting $\phi_{11} \left( x \right) = x_1 x_2$ with $\phi_{10} \left( x \right) = x_1$, \function{monomial} results in $\phi_{21} \left( x \right) = x_1^2 x_2$, while \function{chebyshev}, which implements the Chebyshev basis, yields $\frac{1}{2} \left( \phi_{21}' \left( x \right) + \phi_{01}' \left( x \right) \right)$, where the basis polynomials now correspond to $\phi_{21}' \left( x \right) = \left( 2 x_1^2 - 1 \right) x_2$ and $\phi_{01}' \left( x \right) = x_2$. The \function{<basis>} also contains the necessary information to evaluate the matrix polynomial(s) correctly.
        \item Similarly, the monomial order is important when we need to know the position of a monomial in the polynomial basis (for example, to build the (block) Macaulay matrix or set-up the shift problems). By leaving this computation to a dedicated function that implements the correct monomial order \function{<order>}, the correct position in the chosen monomial order is always obtained. For example, using \function{grinvlex} places the bivariate monomial $x_1^2 x_2$ at position $\num{8}$, while \function{grevlex} yields $\num{9}$.
    \end{itemize}

    By default, \macaulaylab uses the standard monomial basis and graded reverse lexicographic order. 
    The toolbox also contains the pre-implemented Chebyshev basis, graded lexicographic order, graded inverse lexicographic order, and graded negative lexicographic order.
    Users can easily implement and use other polynomial bases or monomial orders.


\section{Functionalities of the toolbox}
    \label{sec:functionalities}

    The previous sections show that the (block) Macaulay matrix can be utilized in several ways, resulting in the different solution approaches implemented in \macaulaylab.
    For example, the choice between using the (right) null space or column space to set up the multiplication maps has a great influence on the results.
    To learn more about the different solution approaches and their options, we once more refer the interested reader to the user manual of \macaulaylab (available at~\cite{vermeersch2025website}).
    In this section, we compare the most important ones (\cref{sec:approaches}).
    However, we first present the database of test problems that is developed alongside the toolbox (\cref{sec:database}), since we will use it to demonstrate \macaulaylab.

    \subsection{Database with examples}
        \label{sec:database}

        Parallel with the development of \macaulaylab, we have also gathered different multivariate polynomial systems and rectangular multiparameter eigenvalue problems in a database, which can be used to test the toolbox's features and act as benchmarks for other software.
        The database is available on the website of \macaulaylab~\cite{vermeersch2025website} and can be downloaded separately.
        It includes a wide variety of problems, which were gathered from our own research, existing repositories (such as \toolbox{Test Database of Polynomial Systems} and \toolbox{Posso Test Suite}), and papers that we encountered during our literature study.
        A reference to the original source is included with every test problem.

        At the time of writing this paper, the database contains $290$ multivariate polynomial systems and $\num{30}$ rectangular multiparameter eigenvalue problem.
        We highlight some entries in \cref{tab:polynomialsystems,tab:eigenvalueproblems}.
        Additional information about the problems is also stored and available in the database, accessible by calling the overloaded function \function{disp}.
        For the problems that we solved via \macaulaylab, we also provide the numerical solutions in a separate file.
        Furthermore, the database includes scripts that illustrate the applications behind the problems and functions to interact with the database.

        Because \macaulaylab is polynomial basis independent (see \cref{sec:basis&order}), the database also contains problems that are given in another polynomial basis. 
        When using such a problem, the \macaulaylab solver recognizes the correct polynomial basis automatically and applies the correct shift and evaluation operations (if the corresponding functions are implemented).
        These problems can be recognized easily in the database, because an identifier for the polynomial basis is appended to the name. 
        For example, \function{problem\_cheb.m} for the Chebyshev polynomial basis.

        \begin{table}
            \sffamily
            \small
            \centering
            \caption{Small selection of multivariate polynomial systems in the database, with some of their key properties: a system has $s$ equations in $m$ variables with maximum total degree equal to $d$, resulting in a total of $m_b$ solutions, of which $m_a$ are affine solutions. The database contains a wide variety of problems. An underscore is used to indicate whether a problem is given in another polynomial basis than the standard monomial basis.}
            \label{tab:polynomialsystems}
            \begin{tabular}{cccccc}
                \toprule 
                \textbf{name} & $\boldsymbol{s}$ & $\boldsymbol{d}$ & $\boldsymbol{m}$ & $\boldsymbol{m_b}$ & $\boldsymbol{m_a}$ \\ 
                \midrule
                \task{noon3} & $\num{3}$ & $\num{3}$ & $\num{3}$ & $\num{27}$ & $\num{21}$ \\
                \task{batselier5} & $\num{3}$ & $\num{12}$ & $\num{3}$ & $\num{1728}$ & $\num{1728}$ \\
                \task{conform} & $\num{3}$ & $\num{4}$ & $\num{3}$ & $\num{64}$ & $\num{16}$ \\
                \task{dreesen10} & $\num{4}$ & $\num{4}$ & $\num{4}$ & $\infty$ & $\num{2}$ \\
                \task{cyclic5} & $\num{5}$ & $\num{7}$ & $\num{5}$ & $\num{120}$ & $\num{70}$ \\
                \task{walsh} & $\num{6}$ & $\num{7}$ & $\num{6}$ & $\infty$ & $\num{7}$ \\
                \task{katsura7} & $\num{8}$ & $\num{2}$ & $\num{8}$ & $\num{128}$ & $\num{128}$ \\
                \task{overdet1} & $\num{8}$ & $\num{4}$ & $\num{4}$ & $\num{10}$ & $\num{10}$ \\
                \midrule
                \task{walsh\_cheb} & $\num{6}$ & $\num{3}$ & $\num{6}$ & $\infty$ & $\num{7}$ \\
                \bottomrule
            \end{tabular}
        \end{table}

        \begin{table}
            \sffamily
            \small
            \centering
            \caption{Small selection of rectangular multiparameter eigenvalue problems in the database, with some of their key properties: an $m$-parameter eigenvalue problem has maximum total degree $d$ and $k \times l$ coefficient matrices, resulting in a total of $m_b$ solutions, of which $m_a$ are affine solutions. The database contains a wide variety of problems. An underscore is used to indicate whether a problem is given in another polynomial basis than the standard monomial basis.}
            \label{tab:eigenvalueproblems}
            \begin{tabular}{ccccccc}
                \toprule 
                \textbf{name} & $\boldsymbol{d}$ & $\boldsymbol{m}$ & $\boldsymbol{k}$ & $\boldsymbol{l}$ & $\boldsymbol{m_b}$ & $\boldsymbol{m_a}$ \\ 
                \midrule
                \task{volkmer} & $\num{1}$ & $\num{2}$ & $\num{12}$ & $\num{6}$ & $\num{6}$ & $\num{6}$ \\
                \task{muhic4} & $\num{1}$ & $\num{2}$ & $\num{18}$ & $\num{9}$ & $\infty$ & $\num{4}$ \\
                \task{alsubaie3} & $\num{1}$ & $\num{3}$ & $\num{5}$ & $\num{3}$ & $\infty$ & $\num{4}$ \\
                \task{h2fom2r3} & $\num{2}$ & $\num{3}$ & $\num{10}$ & $\num{8}$  & $\infty$ & $\num{209}$ \\
                \task{hkp2} & $\num{2}$ & $\num{2}$ & $\num{3}$ & $\num{2}$ & $\num{12}$ & $\num{12}$ \\
                \task{h2f5} & $\num{2}$ & $\num{2}$ & $\num{6}$ & $\num{5}$ & $\infty$ & $\num{17}$ \\
                \task{h2fourdisk} & $\num{2}$ & $\num{4}$ & $\num{10}$ & $\num{7}$ & $\infty$ & $\num{129}$ \\
                \task{cube} & $\num{3}$ & $\num{2}$ & $\num{21}$ & $\num{20}$ & $\num{1890}$ & $\num{1890}$ \\
                \midrule
                \task{wing\_cheb} & $\num{2}$ & $\num{1}$ & $\num{3}$ & $\num{3}$  & $\num{6}$ & $\num{6}$ \\
                \bottomrule
            \end{tabular}
        \end{table}


    \subsection{Different solution approaches in MacaulayLab}
        \label{sec:approaches}

\begin{figure}
            \centering
            \begin{tikzpicture}
    \begin{axis}[
        figurestyle,
        width = 12cm,
        ybar,
        ymin = 0,
        ymax = 10,
        ytick = {1,2,3,4,5,6},
        x tick label style = {text width = 2.25cm, align = center},
        symbolic x coords = {column space column-wise iterative, null space row-wise iterative, null space block-wise iterative, null space block-wise recursive, null space block-wise sparse},
        xtick = data,
        xtick pos = bottom,
        xtick align = inside,
        ylabel = {computation time [\%]},
        xtick style = {white}
    ]

        \addplot[blue, fill = blue] coordinates {
            (column space column-wise iterative,8.0) 
            (null space row-wise iterative,5.8)
            (null space block-wise iterative,1.3)
            (null space block-wise recursive,0.6)
            (null space block-wise sparse,0.4)
        };
        \label{plot:system}

        \addplot[red, fill = red] coordinates {
            (column space column-wise iterative,8.0) 
            (null space row-wise iterative,5.0)
            (null space block-wise iterative,1.1)
            (null space block-wise recursive,0.5)
            (null space block-wise sparse,0.3)
        };
        \label{plot:mep}
    \end{axis}

    \begin{axis}[
        figurestyle,
        width = 12cm,
        ybar,
        ymin = 0,
        ymax = 100,
        ytick = {90,100},
        symbolic x coords = {column space column-wise iterative, null space row-wise iterative, null space block-wise iterative, null space block-wise recursive, null space block-wise sparse},
        xtick = \empty
    ]

        \addplot[blue, fill = blue] coordinates {
            (column space column-wise iterative,100)
            (null space row-wise iterative,0.0)
            (null space block-wise iterative,0.0)
            (null space block-wise recursive,0.0)
            (null space block-wise sparse,0.0)
        };

        \addplot[red, fill = red] coordinates {
            (column space column-wise iterative,100)
            (null space row-wise iterative,0.0)
            (null space block-wise iterative,0.0)
            (null space block-wise recursive,0.0)
            (null space block-wise sparse,0.0)
        };
    \end{axis}

    \draw[white,line width = 0.2cm] (-0.5cm,3cm) -- (12.5cm,3cm);
    \draw[thick] (-0.25cm,2.9cm) -- (0.25cm,2.9cm);
    \draw[thick] (-0.25cm,3.1cm) -- (0.25cm,3.1cm);
    \draw[thick] (11.75cm,2.9cm) -- (12.25cm,2.9cm);
    \draw[thick] (11.75cm,3.1cm) -- (12.25cm,3.1cm);
\end{tikzpicture}
            \ifarxiv \else
                \Description{Different configurations for MacaulayLab are used to solve a multivariate polynomial system and a rectangular multiparameter eigenvalue problem. The options are mentioned below the computation times: the first option is the solution space, the second option is the decision of how the rank checks and shifts in the multiplication maps are performed, and the third option is the chosen enlargement strategy.}
            \fi
            \caption{Different configurations for \macaulaylab are used to solve the \task{noon4} multivariate polynomial system \displaylabel{plot:system} and \task{cube} rectangular multiparameter eigenvalue problem \displaylabel{plot:mep}. The options are mentioned below the computation times: the first option is the solution space, the second option is the decision of how the rank checks and shifts in the multiplication maps are performed, and the third option is the chosen enlargement strategy.}
            \label{fig:options}
        \end{figure}
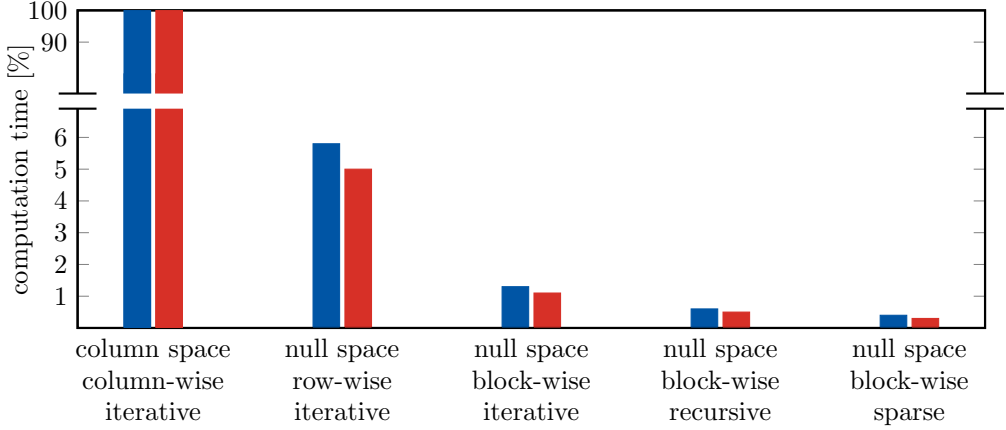

        To demonstrate the different solution approaches of \macaulaylab, we apply the solver to two problems from the database using various configurations. 
        \Cref{fig:options} compares the relative computation times for the \task{noon4} multivariate polynomial system and the \task{cube} rectangular multiparameter eigenvalue problem. 
        The computation time is expressed relative to the slowest configuration in each comparison.
        The first configuration employs the column space approach, while the remaining configurations explore different null space based approaches.

        The first two configurations utilize column-wise and row-wise rank checks combined with column/row-wise shift operations. 
        These configurations are notably slower due to the large number of rank checks required. 
        Rank checks on the columns of the (block) Macaulay matrix are more time-consuming than those on the rows of the null space, as the column vectors are typically longer.

        The recursive and sparse enlargement of the solution space further reduce the computation time. 
        Although the performance gains are modest for the selected two examples, these configurations become significantly more efficient for larger problems. 
        The sparse method offers the additional advantage of eliminating the need to store the, often large, (block) Macaulay matrix in memory.

        Enabling the \function{posdim = true} flag roughly doubles the computation time for both problems. 
        This option introduces numerous unnecessary rank checks (for problems with zero-dimensional solution sets), which results in substantial computational overhead.
        It is, however, a necessary evil when the solution set at infinity is positive-dimensional.

    \section{Comparison with other solvers}
        \label{sec:comparison}

        We compare \macaulaylab with other solvers for multivariate polynomial systems and rectangular multiparameter eigenvalue problems. 
        The test problems that we consider are a subset of our database in \cref{sec:database}.
        We use performance profiles to visualize the comparison.
        For a thorough explanation on performance profiles, we refer the interested reader to~\cite{dolan2002benchmarking}.
        A minimal introduction is given below. 

        For each test problem $i$ in the test set $I$ with $n_i$ problems and each solver $j$ in the comparison set $J$ with $n_j$ solvers, we use $t_{i,j}$ to denote the computation time required to solve problem $i$ by solver $j$.
        The performance of a solver $j$ on a problem $i$ is given by the performance ratio
        \begin{equation}
            r_{i,j} = \frac{t_{i,j}}{\min_{j'} \{t_{i,j'}: j' \in J\}},
        \end{equation}
        which compares the solver $j$ with the best result.
        The performance profile $\rho_j(\tau)$ of a solver $j \in J$,
        \begin{equation}
            \rho_j(\tau) = \frac{1}{n_i} \# \{i \in I: r_{i,j} \leq \tau\}
        \end{equation}
        reveals the probability for the solver $j$ to be within a factor $\tau$ of the best possible performance ratio.
        The performance profile $\rho_j(\tau) : \mathbb{R} \to \left\lbrack0, 1\right\rbrack$ is the (cumulative) distribution function for the performance ratio $r_{i,j}$ and is a nondecreasing, piecewise constant function.
        Notice that the value of $\rho_j(1)$ is the probability that the solver $j$ will be better than the other solvers in $J$.

        \subsection{Other multivariate polynomial system solvers}
            \label{sec:othermultivariatepolynomialsystemsolvers}

\begin{figure}
                \centering
                \begin{tikzpicture}
    \begin{axis}[
        figurestyle,
        xmin=1,
        xmax=9,
        ymin=0,
        ymax=1.0,
        xlabel = {ratio $\tau$},
        ylabel = {performance $\rho_j(\tau)$}
    ]

        \addplot[const plot, bluestyle] table[row sep = \\, col sep = &]{ 
            1 & 0.2968\\
            2 & 0.4552\\
            3 & 0.6087\\
            4 & 0.6087\\
            5 & 0.6232\\
            6 & 0.6232\\
            7 & 0.6232\\
            8 & 0.6812\\
            9 & 0.6812\\
        };
        \label{plot:macaulaylab-system}

        \addplot[const plot, redstyle] table[row sep = \\, col sep = &]{ 
            1 & 0.0045\\
            2 & 0.20174\\
            3 & 0.2309\\
            4 & 0.2309\\
            5 & 0.2354\\
            6 & 0.2899\\
            7 & 0.3478\\
            8 & 0.3478\\
            9 & 0.3478\\
        };
        \label{plot:pnla-svd}

        \addplot[const plot, yellowstyle] table[row sep = \\, col sep = &]{ 
            1 & 0.1439\\
            2 & 0.2074\\
            3 & 0.2119\\
            4 & 0.2119\\
            5 & 0.2119\\
            6 & 0.2464\\
            7 & 0.2464\\
            8 & 0.2464\\
            9 & 0.2609\\
        };
        \label{plot:pnla-qr}

        \addplot[const plot, greenstyle] table[row sep = \\, col sep = &]{ 
            1 & 0.5548\\
            2 & 0.5548\\
            3 & 0.5948\\
            4 & 0.5993\\
            5 & 0.5993\\
            6 & 0.5993\\
            7 & 0.6372\\
            8 & 0.6617\\
            9 & 0.6617\\
        };
        \label{plot:phclab}
    \end{axis}
\end{tikzpicture}%
                \ifarxiv \else
                    \Description{Performance profiles for a subset of 50 multivariate polynomial systems. We run MacaulayLab, PNLA, using the singular value or QR decomposition, and PHClab on this subset.}
                \fi
                \caption{Performance profiles for a subset of $\num{50}$ multivariate polynomial systems. We run \macaulaylab \displaylabel{plot:macaulaylab-system}, \toolbox{PNLA}, using the singular value \displaylabel{plot:pnla-svd} or QR decomposition \displaylabel{plot:pnla-qr}, and \toolbox{PHClab} \displaylabel{plot:phclab} on this subset. In more than half of the problems, \toolbox{PHClab} is the fastest solver, while \macaulaylab has a comparable performance on the test set when we increase the ratio $\tau$. \macaulaylab even manages to solve a slightly larger subset of the problems.}
                \label{fig:performancesystem}
            \end{figure}
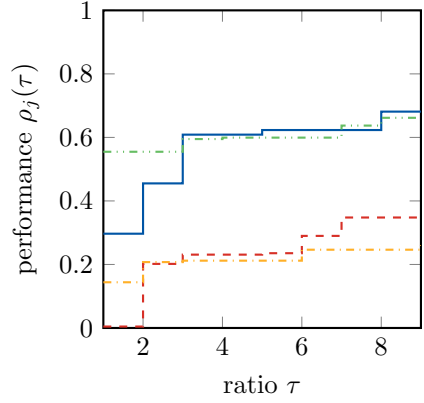

            We compare \macaulaylab with the \toolbox{PNLA} toolbox, which is using the singular value or QR decomposition, \toolbox{PHClab} toolbox (a \matlab interface to \toolbox{PHCpack}), and \toolbox{Maple}.
            The test problems is a random subset of $\num{50}$ multivariate polynomial systems of the database.
            The performance profile is given in \cref{fig:performancesystem}.
            While \toolbox{PHClab} seems to be the fastest solver for a majority of the problems, \macaulaylab has a comparable performance on the test set when we increase the ratio $\tau$. 
            \macaulaylab even manages to solve a slightly larger subset of the problems.
        
        \subsection{Other multiparameter eigenvalue solvers}
            \label{sec:otherrectangularmultiparametereigenvaluesolvers}
        
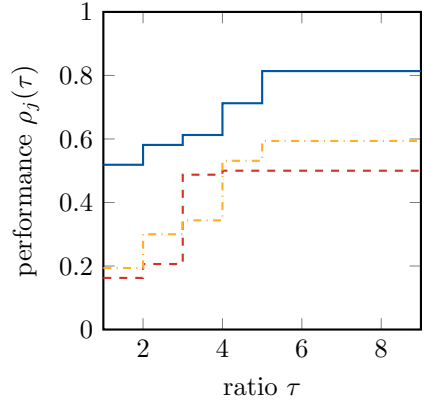
\begin{figure}
                \centering
                \begin{tikzpicture}
    \begin{axis}[
        figurestyle,
        xmin=1,
        xmax=9,
        ymin=0,
        ymax=1.0,
        xlabel = {ratio $\tau$},
        ylabel = {performance $\rho_j(\tau)$}
    ]

        \addplot[const plot, bluestyle] table[row sep = \\, col sep = &]{ 
            1 & 0.51875\\
            2 & 0.58125\\
            3 & 0.6125\\
            4 & 0.7125\\
            5 & 0.81375\\
            6 & 0.81375\\
            7 & 0.81375\\
            8 & 0.81375\\
            9 & 0.81375\\
            10 & 0.84375\\
        };
        \label{plot:macaulaylab-mep}

        \addplot[const plot, redstyle] table[row sep = \\, col sep = &]{ 
            1 & 0.1625\\
            2 & 0.20625\\
            3 & 0.4875\\
            4 & 0.5\\
            5 & 0.5\\
            6 & 0.5\\
            7 & 0.5\\
            8 & 0.5\\
            9 & 0.5\\
            10 & 0.5\\
        };
        \label{plot:multipareig-compress}

        \addplot[const plot, yellowstyle] table[row sep = \\, col sep = &]{ 
            1 & 0.19375\\
            2 & 0.30000\\
            3 & 0.34375\\
            4 & 0.53125\\
            5 & 0.59375\\
            6 & 0.59375\\
            7 & 0.59375\\
            8 & 0.59375\\
            9 & 0.59375\\
            10 & 0.59375\\
        };
        \label{plot:multipareig-sketch}
    \end{axis}
\end{tikzpicture}%
                \ifarxiv \else
                    \Description{Performance profiles for 30 rectangular multiparameter eigenvalue problems. We compare MacaulayLab with the Kronecker matrix approach and randomized sketching approach of MultiParEig. The problem set contains a lot of polynomial problems, which explains the overall better preformance of MacaulayLab.}
                \fi
                \caption{Performance profiles for $\num{30}$ rectangular multiparameter eigenvalue problems. We compare \macaulaylab \displaylabel{plot:macaulaylab-mep} with the Kronecker matrix approach \displaylabel{plot:multipareig-compress} and randomized sketching approach \displaylabel{plot:multipareig-sketch} of \toolbox{MultiParEig}. The problem set contains a lot of polynomial problems, which explains the overall better preformance of \macaulaylab.}
                \label{fig:performancemep}
            \end{figure}

            For rectangular multiparameter eigenvalue problems, the $\num{30}$ problems in the databased are solved by \macaulaylab and \toolbox{MultiParEig}.
            Since \toolbox{MultiParEig} only works with linear problems, all polynomial problems are linearized prior to solving them with the toolbox.
            While the most efficient null space based approach is used for \macaulaylab, two approaches of \toolbox{MultiParEig} are used in the comparison: a compressed operator determinant approach that creates the multiplication maps directly from Kronecker products of the coefficient matrices and a randomized sketching approach that transforms the problem into a coupled square multiparameter eigenvalue system.
            The performance profiles of the three solvers are given in \cref{fig:performancemep}.
            Because the problem set contains a lot of polynomial problems with modestly sized coefficient matrices, \macaulaylab has an overall better performance than the other two approaches that require a linearized version of the problems.
            \macaulaylab is also faster for a high number of scalar parameters, while the operator determinant approach of \toolbox{MultiParEig} obtains better results for problems with a low number of scalar parameters (i.e., two or three) with large coefficient matrices.

\section{Current limitations and future developments}
    \label{sec:limitations&developments}

    Taking inspiration from other toolboxes and recent advances in the literature, it is evident that future releases of \macaulaylab will contain additional features:
    \begin{itemize}
        \item Since the sparse, structured (block) Macaulay matrix constitutes the core of the toolbox, considering the arrangement of the elements in the matrices is an important avenue to further improve the computational capabilities of the code. 
        The recursive and sparse construction of a basis matrix of the null space already do this to a large extent. 
        We will consider the sparsity in the involved matrices and investigate the advantages and disadvantages of working with the QR decomposition instead of the singular value decomposition~\cite{batselier2013numerical}. 
        \item Taking into account the support of the (matrix) polynomials could help to reduce the size of the involved matrices a priori and sounds very promising~\cite{bender2022eigenvalue,bender2022solving}.
        \item Performing rank checks is an important step in the (block) Macaulay matrix approach.
        A lot of care is necessary with these rank decisions.
        In one of the next releases, more support to take the correct rank decisions (i.e., different types of rank decisions, visual aids, etc.) will be added to the toolbox. 
        Related to these rank checks are also the introduction of a block column approach, which would reduce the number of rank checks and improve the robustness of the column space based approach, and the consideration of using projective shifts~\cite{batselier2014null}, which could eliminate the need for rank checks in specific situations (like zero-dimensional solution sets).
        The advantages and disadvantages of these adaptations are prone to additional research.
        \item Instead of computing the eigenvalues of multiple generalized eigenvalue problems independently and using clustering step to combine and refine the solutions, we could use simultaneous triangularization to compute the solutions of the different shift problems faster and more accurate~\cite{he2024randomized}. 
        Combining the different generalized eigenvalue problems in a tensor structure could be another interesting approach to explore~\cite{vanderstukken2021systems}.
        \item Finally, we also want to give the user the possibility to run the solver in an interactive mode, where the user can choose \emph{at run time} which options and approach to use.
        Such an \emph{interactive solver} would be very useful for users who are not familiar with the toolbox or want to learn more about the problem's properties.
    \end{itemize} 

\section{Conclusion}
    \label{sec:conclusion}

    Throughout this paper, we emphasized the versatility and potential of \macaulaylab as a toolbox for solving multivariate polynomial systems and rectangular multiparameter eigenvalue problems. 
    \macaulaylab stands out as a general-purpose multivariate polynomial system solver, capable of dealing with systems in different polynomial bases, overdetermined systems, and systems with positive-dimensional solution sets at infinity.
    By introducing the block Macaulay matrix as natural extension of the (scalar) Macaulay matrix, \macaulaylab becomes one of the first dedicated toolboxex to tackle rectangular multiparameter eigenvalue problems, while maintaining the same flexibility as when solving systems of multivariate polynomial equations. 

    While homotopy continuation computes the solutions faster in many cases, these methods may suffer from ill-conditioning and (only) deal with square systems, leaving a gap that can be filled by the Macaulay matrix algorithms from this (and other) toolbox(es), particularly for overdetermined systems of multivariate polynomial equations. 
    In comparison to other \matlab toolboxes that use the Macaulay matrix, \macaulaylab is a clear improvement: while being built modular and offering many different solution approaches and options, it is computationally more efficient and numerically more robust than its historical predecessors \toolbox{RootFinding} and \toolbox{PNLA}.

    It is important to stress that \macaulaylab is designed to be independent of the polynomial basis and monomial order, which provides an easy way of solving problems that are given in a different representation (without needing to change the representation first). 
    This can be very useful in applications where the results depend on the chosen polynomial basis or monomial order.

    Additionally, the \macaulaylab toolbox also contains a database with many test problems.
    We believe that the inclusion of this comprehensive set of test problems could prove to be highly valuable for its users.

    \section*{Acknowledgments}

        The authors want to thank several colleagues for testing the software packages and providing them with useful comments and suggestions. 
        We acknowledge in that regard the help from Beno\^{i}t Legat, Hans van Rooij, Katrien De Cock, Lukas Vanpoucke, Oscar Mauricio Agudelo, Sarthak De, Sem Viroux, and Sibren Lagauw.

    \bibliographystyle{plain}
    \bibliography{bibliography}
\end{document}